\documentclass[journal]{IEEEtran}
\ifCLASSINFOpdf
\else
\fi
\usepackage{amsfonts}
\usepackage{amsmath,amsfonts,amssymb,epsfig,subfigure,cite,url,multirow,booktabs}
\usepackage{afterpage,algorithm,algorithmic,mathrsfs}
\usepackage{amsthm}

\newtheorem{definition}{\textbf{Definition}}
\newtheorem{lemma}{\textbf{Lemma}}
\newtheorem{theorem}{\textbf{Theorem}}
\newtheorem{remark}{\textbf{Remark}}

\newtheorem{property}{\textbf{Property}}

\usepackage{amsmath}
\begin{document}
%
\title{3-D Deployment of UAV Swarm for Massive MIMO Communications}
%
%
%

\author{Ning Gao,~\IEEEmembership{Member,~IEEE,}~Xiao Li,~\IEEEmembership{Senior Member,~IEEE,}~Shi Jin,~\IEEEmembership{Senior Member,~IEEE,}\\ and Michail Matthaiou,~\IEEEmembership{Senior Member,~IEEE}
\thanks{This work was supported in part by the National Science Foundation
of China under Grants 62001109 \& 61921004, in part by the Key Research and Development Program of Shandong Province under Grant 2020CXGC010108, in part by the China Postdoctoral Science Foundation
under Grants BX20200083 \& 2020M681456, and in part
by the Fundamental Research Funds for the Central
Universities of China under Grant 2242020R20011. This article was presented in part at the ACM MobiArch 2020, London, United Kingdom, September, 2020. (\emph{Corresponding author: Shi Jin.})}
\thanks{N. Gao, X. Li and S. Jin are with the National Mobile Communications
Research Laboratory, Southeast University, Nanjing 210096, China, (e-mail: ninggao@seu.edu.cn; li\_xiao@seu.edu.cn; jinshi@seu.edu.cn).}
\thanks{M. Matthaiou is with the Institute of Electronics, Communications and Information Technology (ECIT), Queen’s University
Belfast, BT3 9DT, U.K., (email:
m.matthaiou@qub.ac.uk).}

}

%
%

\markboth{IEEE JOURNAL ON SELECTED AREAS IN COMMUNICATIONS}%
{Shell \MakeLowercase{\textit{et al.}}: Bare Demo of IEEEtran.cls for IEEE Journals}
%



\maketitle

\begin{abstract}
We consider the uplink transmission between a multi-antenna ground station and an unmanned aerial vehicle (UAV) swarm. The UAVs are assumed as intelligent agents, which can explore their optimal three dimensional (3-D) deployment to maximize the channel capacity of the multiple input multiple output (MIMO) system. Specifically, considering the limitations of each UAV in accessing the global information of the network, we focus on a decentralized control strategy by noting that each UAV in the swarm can only utilize the local information to achieve the optimal 3-D deployment. In this case, the optimization problem can be divided into several optimization sub-problems with respect to the rank function. Due to the non-convex nature of the rank function and the fact that the optimization sub-problems are coupled, the original problem is NP-hard and, thus, cannot be solved with standard convex optimization solvers. Interestingly, we can relax the constraint condition of each sub-problem and solve the optimization problem by a formulated UAVs channel capacity maximization game. We analyze such game according to the designed reward function and the potential function. Then, we discuss the existence of the pure Nash equilibrium in the game. To achieve the best Nash equilibrium of the MIMO system, we develop a decentralized learning algorithm, namely decentralized UAVs channel capacity learning. The details of the algorithm are provided, and then, the convergence, the effectiveness and the computational complexity are analyzed, respectively. Moreover, we give some insightful remarks based on the proofs and the theoretical analysis. Also, extensive simulations illustrate that the developed learning algorithm can achieve a high MIMO channel capacity by optimizing the 3-D UAV swarm deployment with the local information.
\end{abstract}


%
\IEEEpeerreviewmaketitle

\section{Introduction}
\IEEEPARstart{U}{nmanned}
 aerial vehicle (UAV) communication is regarded as a very promising technology in future 6G communication networks. Thanks to its inherent advantages, such as high mobility, deployment flexibility, and strong line-of-sight (LoS) of the air-to-ground channel \cite{7470933}. On one hand, UAVs can be connected to the ground cellular infrastructures as new aerial users to perform their respective missions, which is referred to as cellular-connected UAVs. On the other hand, UAVs can be deployed as new aerial base stations and relays to enhance terrestrial wireless coverage, a paradigm known as UAV-assisted wireless communication.

Several interesting issues arise from the study of flying UAVs in wireless networks, such as trajectory planning \cite{QWUtrajectory,8873597,energyfix}, energy consumption \cite{energyfix,costZeng,YzengJxurotray}, security \cite{8873597}, etc. Particularly, the problem of quasi-stationary UAV deployment, where the locations of UAV are unchanged for the duration of interest to maximize the communication metrics, like throughput, radio coverage radius, etc., has been extensively studied in recent works \cite{8918497}. As an early attempt, the maximum radio coverage of a low-altitude aerial platform has been analyzed with one  dimensional (1-D) altitude optimization \cite{6863654}. Later, considering the UAV-assisted aerial base station, the placement of two dimensional (2-D) aerial base stations and resource allocation were jointly optimized with the goal of maximizing the time-frequency throughput in \cite{8543647}. To further exploit the degrees of spatial freedom, the authors of \cite{7918510} proposed an optimal three dimensional (3-D) placement algorithm to achieve the maximum energy-efficient coverage rate. As a flying relay, the placement of a UAV has been studied to obtain the best trade-off between minimizing
propagation distances to ground terminals and discovering good
propagation conditions \cite{Jchen}. In this context, compared to single UAV communication, UAV swarm communication can support many exciting applications thanks to its enhanced communication reliability, wider coverage area, and resilience to physical attacks. More specifically, in cellular Internet of UAVs, a sensory UAV swarm can be applied to large-scale area monitoring, such as atmospheric and agricultural environmental monitoring. On the battlefield, a UAV swarm network is essential for carrying out the military surveillance tasks and intelligence gathering \cite{8918497}. Likewise, the deployment of UAV swarm can be optimized. In \cite{Jzeng}, the 2-D deployment of multi-UAVs was optimized with the consideration of the line-of-sight (LoS) channel. As an extension of the work \cite{8543647}, joint optimization for resource allocation
and 2-D aerial base station placement were investigated to maximize the downlink sum rate in \cite{8936931}. To maximize the network throughput gain, the multi-UAV 2-D placement was studied in \cite{8675440}. A framework using altitude fixed multi-UAV to optimize the average data rate, while considering the fairness of users, was developed in \cite{Mozaffari}. The UAV-mounted aerial base stations were investigated in \cite{7762053,8760424} with jointly optimizing UAVs 2-D placement and resource allocation. As relay components, the quasi-stationary multi-UAV 2-D placement was studied to maximize the communication reliability in \cite{8116613}. In the context of 3-D space deployment, there are also some works. The authors in \cite{8727504} proposed a multiple UAVs joint 3-D trajectory design and power control strategy via a multi-agent Q-learning algorithm. Based on the sphere packing theory, the optimal 3-D position of the UAVs was studied with jointly considering the coverage area and the flight onboard energy in \cite{MMWS}. Similarly, by leveraging a disk covering model, the reference \cite{8314556} proposed a pattern formation driven 3-D UAV base stations placement, considering different power consumptions of UAVs and the ground user density.

All these works above have provided very useful insights into the deployment of UAVs. However, two crucial issues have never been well considered in the context of optimizing multi-UAV deployment, which are: i) \textit{How does the optimizer access the information of the multi-UAV communication networks if the UAV energy is limited and the remote communication is unreliable?} ii) \textit{How to optimize the multi-UAV deployment if the global information of the networks is not available at the optimizer?} Generally, to accommodate multi-UAVs, a centralized optimization deployment can access all the UAVs of the swarm and keep obtaining the global information at all time, hence, a continuous real-time interaction between each UAV and a computing center is required. For example, many references, i.e., \cite{Jzeng,kang20203d,MMWS,8936931,8314556,8675440,7762053,8760424}, propose to access the global information to perform the optimization placement of multi-UAV via the wireless fronthaul/backhaul link of the optimization control unit. Although a centralized optimization deployment is indeed capable of deploying the UAVs by collecting the information from all parts of the network, it faces two fundamental challenges (\textbf{C1},\textbf{C2}), especially in rural or disaster settings where the remote communication with a computing center might be required:
\begin{itemize}
  \item \textbf{C1}. Since UAVs are often powered by batteries, their airborne time is limited. The centralized optimization deployment entails a large amount of communication overhead to continually schedule and control each UAV, which is not suitable for energy-limited UAV communication.
  \item \textbf{C2}. Due to the unreliable nature of wireless communications, the centralized optimization deployment may be problematic in ultra-reliable low-latency communication. Indeed, this latency may become intolerable for delay-sensitive communications, such as UAV communication and vehicular communication.
\end{itemize}
To overcome the challenges above, developing a decentralized optimization deployment cast a bright light on the multi-UAV deployment problem \cite{8727504,8683875,romero2019noncooperative,Lee2010}. Leaving aside the optimization control centre for the moment, in prior work such as \cite{8727504}, each UAV is required to know the global environment information and the Q-values. However, it is not easy for UAVs to reconstruct the global environment information via the local information exchange, especially when the information is coupled and the memory is limited. Recently, a few works, such as \cite{8683875,romero2019noncooperative,Lee2010}, have investigated the distributed multi-UAV placement. In \cite{8683875}, the backhaul throughput was optimized by adjusting the UAV swarm placement with distributed algorithms. In fact, the proposed distributed algorithms still request one UAV to act as a group leader to coordinate the position of other UAVs in real-time, which is very energy-consuming. The authors in \cite{romero2019noncooperative} proposed a non-cooperative aerial base station placement algorithm leveraging stochastic optimization and machine learning techniques. Although such algorithm does not need aerial base stations to exchange information, it still requires to receive all the packets from the terrestrial infrastructure and to estimate the global information to calculate the network utility function via the real-time feedback information of the control channel. In \cite{Lee2010}, an optimal deployment of multi-UAV was established in wireless airborne communication based on an adaptive hill-climbing type decentralized control algorithm. In that paper, the algorithm includes two modes and needs the throughput among ground nodes, aerial vehicles, as well as an artificial node to construct the cost functions, which induces a considerable latency and algorithm complexity. These observations lead to another challenge:
\begin{itemize}
  \item \textbf{C3}. Most of the decentralized optimization deployments are based on global information acquisition. However, in many emerging scenarios, the global information of the networks may not be available for ability-limited UAVs, such as sensory UAV swarm networks.
\end{itemize}
In a nutshell, most of the existing works address the deployment of UAV swarm by using the centralized optimization deployment or decentralized optimization deployment assuming that the global information is available at all/partial UAVs, which requires a prohibitive amount of resources.

Game theory was first emerged in contemporary economics and quickly found a widespread applications in finance, management science and sociology \cite{MJ1994}. At present, the game theory has been effectively used for wireless resource allocation and optimization in UAV communications \cite{8723552}. For example, in \cite{8478374}, the access selection and resource allocation have been jointly optimized in a UAV swarm assisted communication network by formulating the problem as a dynamic evolutionary game. The N-player normal form games have been proposed to find the load balancing in multiple UAV-assisted wireless communications \cite{gameWIFI}. A bio-inspired and competitive game based flight control of a UAV swarm was studied in \cite{9020811}. The authors in \cite{gamecoordination} utilized the evolutionary game approach to deploy the multi-UAV base stations. They demonstrated that the non-cooperative game is an efficient mathematical tool which can be applied to analyze and optimize the radio communication performance of the UAV swarm.

Different from the existing optimization deployment approaches for multi-UAV, such as \cite{7762053,8760424,8116613,8727504,8683875,romero2019noncooperative,Lee2010},
in this paper, we propose a novel decentralized deployment algorithm which is based on the non-cooperative game. Based on such an algorithm, the coordinated behavior for each UAV is locally controlled, and the individual UAV deployment is only based on the local information from the neighbors, and there is no need to coordinate via group leadership or global information. Thus, the aforementioned challenges \textbf{C1}-\textbf{C3} can be effectively addressed. Particularly, focusing on the effectiveness of communication at high signal-to-noise ratio (SNR), we propose a decentralized learning algorithm for the UAV swarm deployment, to maximize the channel capacity of the multiple input multiple output (MIMO) system. The developed learning algorithm, namely decentralized
UAVs channel capacity learning, is based on the potential game \cite{4814554}, which has been successfully applied in distributed wireless resource allocations \cite{6086561,8654694,5506285}. By using the developed learning algorithm, each UAV only needs to exchange local information with its neighbors and executes the local real-time computation to explore its optimal 3-D deployment, hereby maximizing the channel capacity of the MIMO system. We show that the MIMO system can achieve a high channel capacity via the developed learning algorithm in a finite number of iterations. To the best of our knowledge, the decentralized 3-D deployment of UAV swarm by considering only the local information in massive MIMO communication has not been well studied yet. Specifically, the main contributions of this paper are summarized as follows:
\begin{itemize}
  \item We propose a paradigm for massive MIMO communication with a 3-D deployed UAV swarm. We formulate the MIMO channel capacity maximization problem into several rank function maximization sub-problems. To make the optimization problem more tractable, we relax the constraint condition requiring the MIMO channel matrix to be orthogonal.
  \item To solve the optimization problem only with the local information, we model the UAV swarm communication as an undirected sparse connected graph and solve the optimization problem by the formulated UAVs channel capacity maximization game.
      The solution with respect to the pure Nash equilibrium of the game is analyzed.
  \item To obtain the global optimization solution with respect to  the best Nash equilibrium of the game, we develop the decentralized UAVs channel capacity learning. The details of the proposed algorithm are provided and the performances in terms of convergence and effectiveness are analyzed. In addition, we give some insightful remarks based on theoretical analysis and numerical simulations.
\end{itemize}

The rest of the paper is organized as follows: In Section \ref{sec:2}, we present the system model and problem formulation. In Section \ref{sec:3}, we transform the optimization problem to a series of distributed optimization sub-problems and propose potential game to solve them. In Section \ref{sec:4}, to achieve the best Nash equilibrium of the game, we develop the decentralized learning algorithm. Simulations are presented in Section \ref{sec:5} and conclusion is given in Section \ref{sec:6}. A list of commonly used notations throughout this paper is summarized in Table \ref{Tab:notations}.
\begin{table}[h]
\caption{List of notations}
\label{Tab:notations}
\centering
\begin{tabular}{c|c}
  \hline
 \textbf{Symbol} & \textbf{Definition} \\
  \hline
  $\mathbf{A}$ & Matrix \\
  $\mathbf{a}$ & Vector \\
  $(\cdot)^H$ & Hermitian transpose of the matrix \\
  $(\cdot)^\dag$ & Hermitian transpose of the vector \\
  $\mathbb{C}^{N\times M}$ & Dimension of the complex matrix \\
  $\text{rank}(\cdot)$ & Matrix rank operator \\
  $\text{tr}(\cdot)$ & Matrix trace operator \\
  $\|\cdot\|$& 2-norm operator  \\
  $|\cdot|$& Size operator  \\
  $a\rightarrow a'\rightarrow\cdots$ & State/action transition \\
  \hline
\end{tabular}
\end{table}
\section{system model and problem formulation}\label{sec:2}
In this section, we first give the system model, then, we formulate the optimization problem.
\subsection{System Model}
\begin{figure}[!ht]
 \centering
  \includegraphics[width=8.5cm]{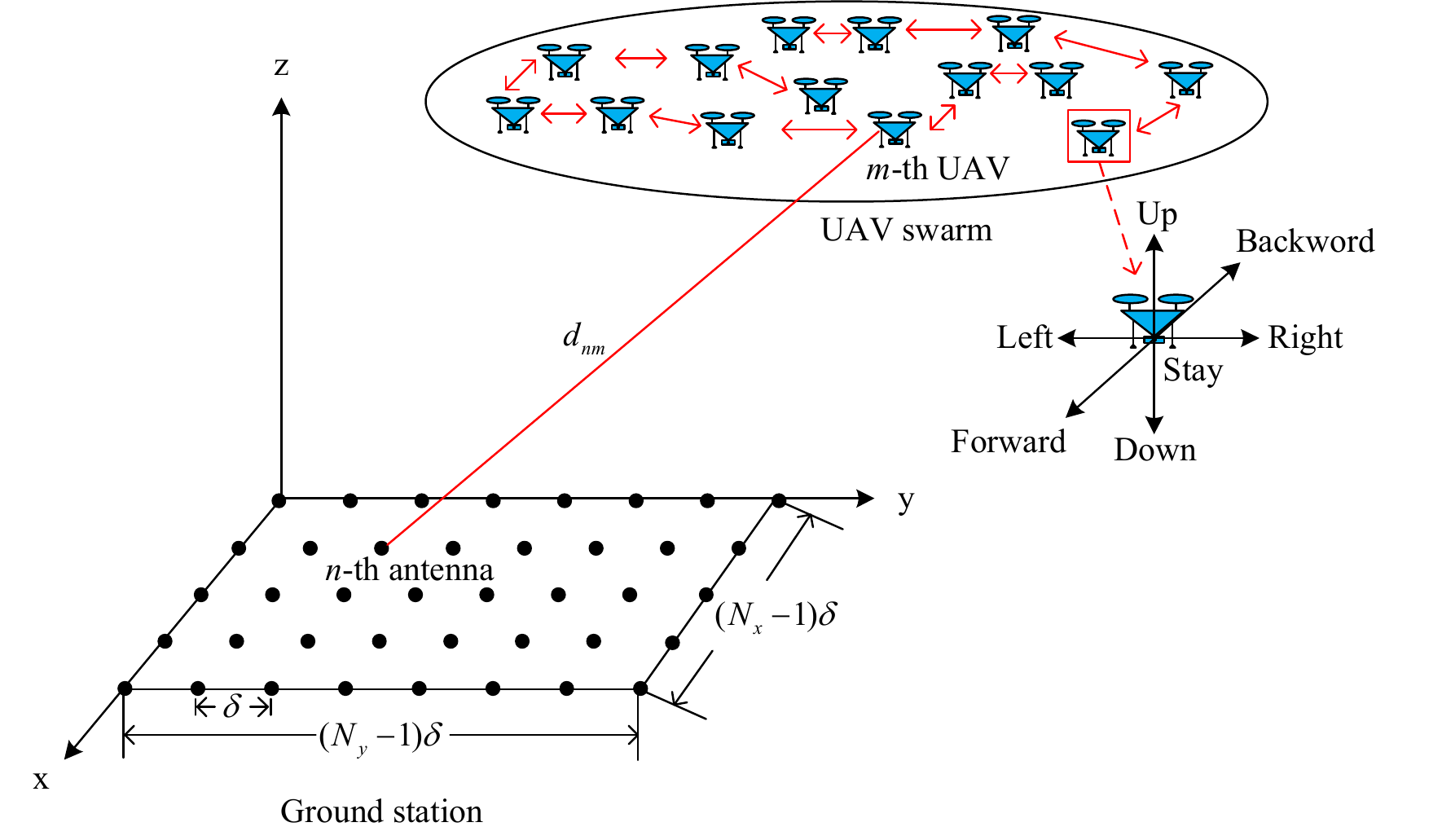}
  \caption{The uplink communication between a URA and a UAV swarm. The UAV swarm is deployed in a three-dimension space and the URA of the GS is deployed in a two-dimension plane.}
  \label{Fig:model}
\end{figure}
We consider the uplink transmission between a ground station (GS) and a UAV swarm with $M$ rotary-wing UAVs, which is shown in Fig. \ref{Fig:model}. The GS is equipped with a uniform rectangular array (URA) with $N$ antennas, and each UAV in the swarm is equipped with a single antenna. Along the plane of the URA, we set an orthonormal coordinate system with unit basis vectors $\mathbf{x}$, $\mathbf{y}$, and $\mathbf{z}$, where $N_x$ antennas are on the $x$-axis and $N_y$ antennas are on the $y$-axis. Then, the total number of antennas can be denoted as $N=N_xN_y$. Assuming the spacing between each antenna is $\delta$, the position of the $n$-th antenna can be denoted as $(x_n,y_n,z_n)=((n_x-1)\delta),(n_y-1)\delta),0)$, where $n_x\in\{1,2,\ldots,N_x\}$, $n_y\in\{1,2,\ldots,N_y\}$ and $n=(n_y-1)N_x+n_x$. Similarly, the position of the $m$-th UAV can be denoted as $(x^m,y^m,z^m)$ with $m\in\{1,2,\ldots,M\}$. We assume that the position of the URA is fixed with the height dimension ignored, while, in each time slot, the UAVs in the swarm are mobile at constant velocities $\nu_c$ or motionless. Specifically, in one time slot, the $m$-th UAV can select a position from seven directions, including stay, up, down, left, right, forward, backward. We map the moving directions as $\text{stay}\rightarrow(0,0,0)$, $\text{up}\rightarrow(0,0,1)$, $\text{down}\rightarrow(0,0,-1)$, $\text{left}\rightarrow(0,-1,0)$, $\text{right}\rightarrow(0,1,0)$, $\text{forward}\rightarrow(1,0,0)$, $\text{backward}\rightarrow(-1,0,0)$ and denote the mapping set as the action space
\begin{eqnarray}
\mathcal{A}_m=\{(0,0,0),(0,0,1),(0,0,-1),(0,-1,0),(0,1,0),\nonumber\\
(1,0,0), (-1,0,0)\},\nonumber
\end{eqnarray}
which represents the finite set of the discrete spatial positions that the $m$-th UAV can select from\footnote{It should be noted that the action space can be easily extended to a larger one by quantizing the horizontal moving direction $\theta\in[0,2\pi)$, the  horizontal moving direction $\varphi\in[-\frac{\pi}{2},\frac{\pi}{2}]$ and the moving speed $\nu\in[0,\nu_c]$.}.
The UAVs are considered as intelligent agents, who can learn their optimal position selection strategy via environmental interaction. In one time slot, the $m$-th UAV senses the communication environment, chooses an action $a_m\in\mathcal{A}_m$ to determine the position which can maximize its reward, i.e., the individual channel capacity.

We assume that the UAVs can communicate with their neighbors and that the inter-UAV communication link is sparse connectivity and noise-free. The downlink channel between URA and each UAV is a typical air-to-ground channel, which contains strong LoS, reflected nonline-of-sight (NLoS), and small-scale fading. In general, the influence of small-scale fading is smaller than LoS and NLoS \cite{1683399} and the probability of NLoS can be neglected since the UAV swarm has a higher altitude than the building surroundings. Therefore, we use the strong LoS approach to model the air-to-ground channel. The channel between the $n$-th antenna of the URA and the $m$-th UAV is denoted as
\begin{eqnarray}
\label{eq:channel}
h_{mn}=\eta_{mn} e^{-j2\pi\frac{d_{mn}}{\lambda}},~~~~~~~~~~~~~~~~~~~~~~~~~~~~~~~~~~~~~  \nonumber\\
\forall m\in \{1,2,\ldots,M\}, n \in \{1,2,\ldots,N\},
\end{eqnarray}
where the path loss coefficient is denoted as $\eta_{mn}=\frac{\lambda}{4\pi d_{mn}}$, the distance between the antenna $n$ and the UAV $m$ is $d_{mn}=\sqrt{(x^m-x_n)^2+(y^m-y_n)^2+(z^m)^2}$, and the wavelength of the signal is $\lambda$. As the UAVs are placed far from the URA and move in a small space, the relative differences in path loss are negligible, we can normalize the uplink MIMO channel with respect to the distance and represent it as the matrix
\begin{eqnarray}
\label{eq:channelMIMO}
\mathbf{H}=\left[
             \begin{array}{cccc}
               h_{11} & h_{12} & \cdots & h_{1M} \\
               h_{21} & h_{22} & \cdots & h_{2M} \\
               \vdots & \vdots & \ddots & \vdots \\
               h_{N1} & h_{M2} & \cdots & h_{NM} \\
             \end{array}
           \right]\in \mathbb{C}^{N\times M}.
\end{eqnarray}
\subsection{Problem Formulation}
The goal of the optimization problem is to maximize the channel capacity of the MIMO system. For simplicity of exposition, we here consider $N=M$. We can express the receive signal of the GS as
\begin{eqnarray}
\label{eq:receive}
\mathbf{r}=\sqrt{E_s}\mathbf{H}\mathbf{s}+\mathbf{n},
\end{eqnarray}
where $\sqrt{E_s}$ is the transmit power, $\mathbf{s}$ is the transmit signal of the UAV swarm and $\mathbf{n}$ is the noise following zero-mean
complex Gaussian distribution with covariance $\mathbf{I}_NN_0$. Assuming no channel state information on the transmit side, the most meaningful strategy is to consider that the transmitted signal $\mathbf{s}$ composed of $M$ statistically independent equal power components\footnote{Since the UAVs in the swarm are homogeneous, the transmitted signal power of each UAV is equal.}. Then, at the receiver side, i.e., URA, the instantaneous capacity of the MIMO channel can be expressed as \cite{2004Channel}
\begin{eqnarray}
\label{eq:capcaity}
C=\log_2\bigg(\det[\mathbf{I}_N+\frac{\rho}{N} \mathbf{H}\mathbf{H}^H]\bigg),
\end{eqnarray}
where $\rho=E_s/N_0$ is the SNR at each receiving antenna.
Due to the high correlation of the air-to-ground LoS channel, the LoS correlation matrix $\mathbf{R}=\mathbf{H}\mathbf{H}^H$ can be rank-deficient.
Then, based on the singular value decomposition of $\mathbf{R}$, the MIMO channel capacity can be rewritten as
\begin{eqnarray}
\label{eq:capcaity1}
C=\sum_{q=1}^p\log_2\bigg(1+\frac{\rho}{p}\lambda_q\bigg),
\end{eqnarray}
where $\lambda_q$ is the $q$-th eigenvalue of $\mathbf{R}$ and $p\leq N$.
By Jensen's inequality,
\begin{eqnarray}
\label{eq:capcaity1Jen}
\frac{1}{p}\sum_{q=1}^p\log_2\bigg(1+\frac{\rho}{p}\lambda_q\bigg)\leq \log_2\bigg(1+\frac{\rho}{p^2}\bigg(\sum_{q=1}^p\lambda_q\bigg)\bigg),
\end{eqnarray}
we can approximately get
\begin{eqnarray}
\label{eq:capcaity1JenAPP}
C\approx p\log_2\bigg(1+\frac{\rho}{p^2}\text{tr}(\mathbf{R})\bigg),
\end{eqnarray}
where the term $\sum_{q=1}^p\lambda_q=\text{tr}(\mathbf{R})$.
By parameterizing the rank $p=N^\alpha, \alpha\in[0,1]$ and $\text{tr}(\mathbf{R})=N^\gamma, \gamma\in[0,2]$, (\ref{eq:capcaity1JenAPP}) can be rewritten as
\begin{eqnarray}
\label{eq:capcaity2}
C\approx N^\alpha\log_2\bigg(1+\frac{\rho N^\gamma}{p^2}\bigg).
\end{eqnarray}
According to (\ref{eq:capcaity2}), by computing the first and the second order derivatives with respect to $p$, it can be shown that the optimal rank is a function of $\rho$, say $p^*(\rho)$, and the MIMO channel matrix $\mathbf{H}$ should be full rank to achieve the maximum channel capacity at high SNRs\footnote{Rank-1 is optimal at low SNRs, and a multi-UAV beamforming regime should be applied. However, in this paper, we only consider the high SNRs case for which rank-$N$ is optimal \cite{5456446}, and, thus, we focus on a multi-UAV multiplexing regime via 3-D UAV swarm deployment.}, i.e., $p^*(\rho)=N$ \cite{5456446}. Considering the high SNR regime, we can formulate the channel capacity maximization problem as
\begin{eqnarray}
\mathrm{P}0:~~~~\max~\text{rank}(\mathbf{H}),
\end{eqnarray}
which is to maximize the rank of $\mathbf{H}$.
Nevertheless, given their limited sensory ability and onboard energy, it is extremely challenging for each UAV to obtain the global information of the network, i.e., all the position information of the other UAVs. Without such knowledge, it is difficult for each UAV to construct the MIMO channel $\mathbf{H}$ and directly maximize the matrix rank as given in $\mathrm{P}0$. In other words, the MIMO channel matrix $\mathbf{H}$ is unknown to each UAV. Thus, we consider the 3-D deployment for UAVs with a decentralized control strategy. In particular, by exchanging the local information and adjusting the 3-D deployment, each UAV forms a MIMO channel $\mathbf{H}_m,\forall m,\in \{1,2,\ldots,M\}$ with its neighbors and attempts to maximize the channel capacity of such MIMO channel, hereby maximizing the channel capacity of the MIMO system. Thus, the optimization problem for UAV $m$ can be formulated as
\begin{eqnarray}
\mathrm{P}1\text{-m}:~~~~\max~\text{rank}(\mathbf{H}_m),
\end{eqnarray}
where the MIMO channel $\mathbf{H}_m$ is constructed by the LoS channels of the $m$-th UAV to the URA and of its neighbors to the URA, and the dimensions of $\mathbf{H}_m$ are dependent on the number of neighbors of the $m$-th UAV. To this end, the total optimization problem of the MIMO system can be given by
\begin{eqnarray}
\mathrm{P}1:~~~~\max~\text{rank}(\mathbf{H}_m), ~\forall m\in\{1,2,\ldots,M\}.
\end{eqnarray}
\section{Optimization Problem Transformation and Game Analysis}\label{sec:3}
The optimization problem $\mathrm{P}1$ reveals that each UAV has to maximize the rank of $\mathbf{H}_m$ which is jointly formed with its neighbors. However, the rank function is non-convex \cite{Rockafellar1996Convex} and the ranks of any two MIMO channel matrixes i.e., $\mathbf{H}_m,\mathbf{H}_{m+1}$, are coupled with the position selection of their common neighbors; thus, the problem $\mathrm{P}1$ is NP-hard. To solve these challenges, we transform the optimization problem into several optimization sub-problems and solve them as a game.
\subsection{Optimization sub-problem}
Let $\mathbf{H}_m=[\mathbf{h}_1,\ldots,\mathbf{h}_i,\mathbf{h}_m]$, where $\mathbf{h}_i\in \mathbb{C}^{N\times 1}$ is the channel from the $i$-th neighbor of the $m$-th UAV to the GS, and $\mathbf{h}_m\in\mathbb{C}^{N\times 1}$ is the channel from the $m$-th UAV to the GS. For the non-convex rank function, we relax the constraint condition to the best-case that the MIMO channel $\mathbf{H}_m$ has orthogonal columns. It is clear that the relaxation will not change the nature of the optimization problem. Then, the optimization problem in $\mathrm{P}1\text{-m}$ can be transformed to
\begin{eqnarray}
\label{eq:pm2-1}
\mathrm{P}2\text{-m}:&&\min_{\substack{d_{mn}\\n\in\{1,2,\cdots,N\}}}~\|\mathbf{h}_k^\dag \mathbf{h}_l\|,\nonumber\\
&&\text{s.t.}~\forall k,l\in \{1,\ldots,i,m\},k\neq l.
\end{eqnarray}
Then, the term $\|\mathbf{h}_k^\dag \mathbf{h}_l\|$ in problem $\mathrm{P}2\text{-m}$ can be rewritten as
\begin{eqnarray}
\label{eq:moptimal}
\|\mathbf{h}_k^\dag \mathbf{h}_l\|=\|g^{m_{th}}_{kl}\|=\big\|\sum_{n \in\{1,2,\ldots,N\}}e^{-j2\pi\frac{d_{nl}-d_{nk}}{\lambda}}\big\|,~\nonumber\\
\forall k,l\in \{1,\ldots,i,m\},k\neq l,~~\nonumber
\end{eqnarray}
where the superscript $m_{th}$ represents the term for the $m$-th UAV.
In this case, the optimization problem $\mathrm{P}1$ can be formulated as $M$ optimization sub-problems, which can be represented by
\begin{eqnarray}
\label{eq:optimal1}
\mathrm{P}2:\min_{\substack{d_{mn}\\m\in\{1,2,\cdots,M\}\\ n\in\{1,2,\cdots,N\}}} \sum_{m=1}^M \underbrace{\sum_{\forall k,l,k\neq l}\|g^{m_{th}}_{kl}\|}_{\text{The~m-th~UAV}}.
\end{eqnarray}

From (\ref{eq:pm2-1}), we can obtain that the action change of the neighbors can effect the optimal solution of the corresponding UAV. In other words, the optimal solutions for all UAVs are coupled with the actions of their common neighbors. Thus, it is difficult for each UAV to obtain its optimal solution with the local information, simultaneously. Next, we formulate the control strategy selection process as a game and design the special objective functions which allow each UAV to obtain the optimal solution, while guaranteeing the maximum channel capacity of the MIMO system.
\subsection{Game played on graph}
The inter-UAV communication can be formulated as an undirected sparse connected graph $G=(V,E)$ with vertex set $V=\{1,2,\ldots,M\}$ and edge set $E$, which denote the set of UAVs and inter-UAV communication links, respectively. Then, the neighbor set of the $m$-th UAV is defined by
\begin{eqnarray}
\label{eq:neighborset}
\mathcal{N}_m=\{i\in V|(m,i)\in E\}.
\end{eqnarray}
Let $\mathcal{N}_m=\{1,\ldots,i\}$ be the neighbor set of the $m$-th UAV, then, the $m$-th UAV has degree $D_m=|\mathcal{N}_m|$, representing the number of communication links for the $m$-th UAV. In terms of multi-agent learning, each agent regards the other agents as a part of the environment and interacts with such environment to obtain the corresponding rewards, and then determines the optimal control strategy. To obtain the environment information, the neighbor UAVs propagate the local information with each other via the edge $e\in E$ on the graph. To put things in a mathematical context, define a measurement space $(\Omega,\mathcal{F},P)$ with filtration $\{\mathcal{F}^t\}$, decision processes for each $t$ are adapted to $\mathcal{F}^t$, then the global information can be denoted by a $\sigma$-algebra with a collection $\mathcal{M}$ of random objects, i.e., $\mathcal{F}^t=\sigma(\mathcal{M})$. Then, we have
\begin{eqnarray}
\label{eq:informationsum}
\mathcal{F}^t=\bigvee_{m=1}^M\mathcal{F}^t_m,
\end{eqnarray}
where $\mathcal{F}^t_m$ is the local information of the $m$-th UAV, while the symbol $\bigvee$ denotes the ``join" of the $\sigma$-algebra.
\begin{remark}
\label{remark:1}
Here, the key difference between the centralized
computing control and the decentralized
computing control is on the global information $\mathcal{F}^t$ and the local information $\mathcal{F}^t_m$. The former includes the measured random objects for all UAVs at all time, while the latter only has the measured random objects of the $m$-th UAV and its neighbors. In fact, the local information is strictly the subset of the global information, i.e., $\mathcal{F}^t_m \subset \mathcal{F}^t$ for any $m$ and $t$, if the inter-UAV communication link with sparse connectivity.
\end{remark}
In game theory, a game $\mathcal{G}=\{V,\{\mathcal{A}_m\}_{m\in V}, \{R_m\}_{m\in V}\}$ is defined by its three components:
\begin{itemize}
  \item $V=\{1,2,\ldots,M\}$ is the set of UAVs.
  \item $\mathcal{A}_m$ is the set of all actions that the $m$-th UAV has, i.e., the position selection in seven directions. An action profile $a\in \mathcal{A}$ denotes the collection of actions of all the UAVs, where $\mathcal{A}=\mathcal{A}_1\times\cdots\times\mathcal{A}_M$ is the action space of all the UAVs.
  \item $R_m$ is the reward of the $m$-th UAV.
\end{itemize}
The Nash equilibrium is one of the most important concepts in a game, which can be defined as follows.
\begin{definition}
\label{eq:NE}
For a game $\mathcal{G}=\{V,\{\mathcal{A}_m\}_{m\in V}, \{R_m\}_{m\in V}\}$, an action profile $a^*$ is represented as a pure Nash equilibrium of the game, if and only if
\begin{eqnarray}
\label{eq:defNE}
R_m(a^*_m,a^*_{-m})\geq R_m(a_m,a^*_{-m}),\nonumber~~~~~~~~~~~~~~~~~~\\
\forall a_m\in \{\mathcal{A}_m/\ a^*_m\}, \forall m\in V,
\end{eqnarray}
where $\{\mathcal{A}_m/\ a^*_m\}$ means that $a^*_m$ is excluded from $\mathcal{A}_m$ and $a^*_{-m}\in\mathcal{A}_{-m}$ denotes the optimal action profile of all the UAVs except $m$.
\end{definition}
The game $\mathcal{G}=\{V,\{\mathcal{A}_m\}_{m\in V}, \{R_m\}_{m\in V}\}$ has reached a pure Nash equilibrium if and only if no UAV has the motivation to unilaterally change its action. Based on (\ref{eq:optimal1}), we can observe that each UAV can only learn its optimal control strategy without the consideration of the optimal solutions of all UAVs, which is typical in a non-cooperative game. Meanwhile, for the formulated optimization problem, the reward of each UAV should be appropriately aligned to the global reward to solve the rank coupling problem. Therefore, we propose a UAVs channel capacity maximization game based on the non-cooperative potential game. The reward function of a UAV should include its own reward and the aggregate reward of its neighbors. In such a setting, the local information $\mathcal{F}_m^t, \forall m\in V$ in current time slot $t$ can be denoted as
\begin{eqnarray}
\label{info}
\mathcal{F}_m^t=\sigma_m\bigg\{(x^{m,t},y^{m,t},z^{m,t}),(x^{\mathcal{N}_m,t},y^{\mathcal{N}_m,t},z^{\mathcal{N}_m,t})\bigg\},
\end{eqnarray}
and the reward function of the $m$-th UAV can be given by
\begin{eqnarray}
\label{eq:immR}
R_m(a_m,a_{\mathcal{N}_m})=r_m(a_m,a_{\mathcal{N}_m})+\sum_{i\in\mathcal{N}_m}r_i(a_i,a_{\mathcal{N}_i}),
\end{eqnarray}
where $r_m(a_m,a_{\mathcal{N}_m})=-\sum_{\forall k,l,k\neq l}\|g^{m_{th}}_{kl}\|$ and $r_i(a_i,a_{\mathcal{N}_i})=-\sum$ $_{\forall k,l,k\neq l}\|g^{i_{th}}_{kl}\|$, respectively. The symbol $a_{\mathcal{N}_m}\in \mathcal{A}_{\mathcal{N}_m}$ denotes the action profile of the neighbors of the $m$-th UAV. The local game $\mathcal{G}_m$ for the $m$-th UAV can be expressed as
\begin{eqnarray}
\label{eq:localgame}
\mathcal{G}_m:\max_{a_m\in \mathcal{A}_m}R_m(a_m,a_{\mathcal{N}_m}), \forall m\in V,
\end{eqnarray}
where $\mathcal{G}=\mathcal{G}_1\times\cdots\times\mathcal{G}_m\cdots\times\mathcal{G}_M$.
We consider the following potential function, which is defined by
\begin{eqnarray}
\label{eq:potential}
\phi(a_m,a_{-m})=\sum_{m\in V}r_m(a_m,a_{\mathcal{N}_m}).
\end{eqnarray}
The potential function obtains the value $0$ if and only if $r_m=0, \forall m \in V$, which means that both the potential function and the reward of each UAV are maximum.
\begin{theorem}
For the UAVs channel capacity maximization game, the potential function has one and only one solution, $r_m=0, \forall m\in V$, which can make the MIMO system reach the optimal solution.
\end{theorem}

\begin{IEEEproof}
As per (\ref{eq:immR}), we find that each term of (\ref{eq:potential}) is less than or equal to 0, i.e., $r_m(a_m,a_{\mathcal{N}_m})\leq 0, \forall m\in V$, thus, $\phi(a)\leq0$.
Let $\phi(a)=0$, and then we have
\begin{eqnarray}
\label{eq:potential0}
r_1(a_1,a_{\mathcal{N}_1})+\cdots+r_m(a_m,a_{\mathcal{N}_m})+\cdots~~~~~~~~~~~~~~\nonumber\\
+r_M(a_M,a_{\mathcal{N}_M})=0.
\end{eqnarray}
\begin{figure*}
\begin{eqnarray}
~~\phi(a_m,a_{-m})=
-\sum_{\forall k,l\in\{\mathcal{N}_1,1\},k\neq l}\bigg\|\sum_{n\in\{1,2,\cdots,N\}}e^{-j2\pi\frac{d_{nl}-d_{nk}}{\lambda}}\bigg\|-\cdots
-\sum_{\forall k,l\in\{\mathcal{N}_m,m\},k\neq l}\bigg\|\sum_{n\in\{1,2,\cdots,N\}}e^{-j2\pi\frac{d_{nl}-d_{nk}}{\lambda}}\bigg\|\nonumber\\
-\cdots-\sum_{\forall k,l\in\{\mathcal{N}_M,M\},k\neq l}\bigg\|\sum_{n\in\{1,2,\cdots,N\}}e^{-j2\pi\frac{d_{nl}-d_{nk}}{\lambda}}\bigg\|.
\label{eq:consensus}
\end{eqnarray}
\hrulefill
\end{figure*}
Substituting (\ref{eq:moptimal}) into (\ref{eq:potential}), we can get (\ref{eq:consensus}) which is shown at the top of the next page. After some
algebraic computations, we can find that if $\phi(a)=0$ holds, each term of (\ref{eq:consensus}) should be 0, i.e.,
\begin{eqnarray}
\label{eq:special}
\left\{
  \begin{array}{ll}
    \sum_{\forall k,l\in\{\mathcal{N}_1,1\},k\neq l}\|g^{1_{st}}_{kl}\|=0, & \hbox{} \\
    ~~~~~~~~\vdots & \hbox{}\\
    \sum_{\forall k,l\in\{\mathcal{N}_m,m\},k\neq l}\|g^{m_{th}}_{kl}\|=0, & \hbox{} \\
    ~~~~~~~~\vdots & \hbox{}\\
    \sum_{\forall k,l\in\{\mathcal{N}_M,M\},k\neq l}\|g^{M_{th}}_{kl}\|=0. & \hbox{}
  \end{array}
\right.\nonumber
\end{eqnarray}
Therefore, $r_m=0, \forall m\in V$ is an unique solution of $\phi(a)=0$, which makes the MIMO system reach the optimal solution. These conditions can be satisfied when and only when each MIMO channel $\mathbf{H}_m$ is orthogonal, which corresponds to the maximum channel capacity of the MIMO system, and the proof is completed.

\end{IEEEproof}
\begin{remark}
The global optimal solution of the UAV channel capacity game can be fully characterized by the potential function (\ref{eq:potential}) which represents the sum of the UAV rewards. The reason is that the reward function of each UAV and the potential function are perfectly aligned. In this case, the reward function and the potential function can be maximized at the same time, i.e., $R_m(a_m,a_{\mathcal{N}_m})=0, \forall m\in V$ and $\phi(a)=0$. Intuitively, the action profile that maximizes the potential function is a pure Nash equilibrium. However, the strict mathematical proof is needed to show that the proposed UAV channel capacity game is an exact potential game and has a pure Nash equilibrium.
\end{remark}
\subsection{Analysis of game}
\begin{theorem}
\label{eq:theorem2}
The UAV channel capacity maximization game is an exact potential game which has at least one pure Nash equilibrium.
\end{theorem}
\begin{IEEEproof}
Assuming that an arbitrary UAV $m$ unilaterally changes its position selection from $a_m$ to $a'_m$, then the change in the reward of the $m$-th UAV by this unilateral change can be given by
\begin{eqnarray}
\label{eq:change}
R_m(a'_m,a_{\mathcal{N}_m})-R_m(a_m,a_{\mathcal{N}_m}).
\end{eqnarray}
 Substituting (\ref{eq:immR}) into (\ref{eq:change}), we can get
\begin{eqnarray}
\label{eq:change1}
R_m(a'_m,a_{\mathcal{N}_m})-R_m(a_m,a_{\mathcal{N}_m})=
r_m(a'_m,a_{\mathcal{N}_m})~~~~~~~\nonumber\\
-r_m(a_m,a_{\mathcal{N}_m})+\sum_{i\in\mathcal{N}_m}r_i(a_i,a'_{\mathcal{N}_i})
-\sum_{i\in\mathcal{N}_m}r_i(a_i,a_{\mathcal{N}_i}),
\end{eqnarray}
where $r_m(a'_m,a_{\mathcal{N}_m})$ and $r_i(a_i,a'_{\mathcal{N}_i})$ represent the individual reward of the $m$-th UAV and one of its neighbor after unilaterally changing the action from $a_m$ to $a'_m$.
On the other hand, the change of the potential function by switching the action of the $m$-th UAV from $a_m$ to $a'_m$, provided that all the other actions of that UAV remain unchanged, is
\begin{eqnarray}
\label{eq:change2}
\phi(a'_m,a_{-m})-\phi(a_m,a_{-m})
=r_m(a'_m,a_{\mathcal{N}_m})-r_m(a_m,a_{\mathcal{N}_m})\nonumber\\
+\sum_{i\in\mathcal{N}_m}r_i(a_i,a'_{\mathcal{N}_i})
-\sum_{i\in\mathcal{N}_m}r_i(a_i,a_{\mathcal{N}_i})\nonumber\\
+\sum_{i\in\{ V/\ \mathcal{N}_m\},i\neq m}r_i(a_i,a'_{\mathcal{N}_i})
-\sum_{i\in\{ V/\ \mathcal{N}_m\},i\neq m}r_i(a_i,a_{\mathcal{N}_i}).\nonumber
\end{eqnarray}
Since the action of the $m$-th UAV only affects the rewards of its neighbors, we have
\begin{eqnarray}
\label{eq:change3}
r_i(a_i,a'_{\mathcal{N}_i})
-r_i(a_i,a_{\mathcal{N}_i})=0, \forall i\in\{ V/\ \mathcal{N}_m\},i\neq m.
\end{eqnarray}
Then, we can get
\begin{eqnarray}
\phi(a'_m,a_{-m})-\phi(a_m,a_{-m})=~~~~~~~~~~~~~~~~~~~~~~~~~\nonumber\\
R_m(a'_m,a_{\mathcal{N}_m})-R_m(a_m,a_{\mathcal{N}_m}),
\end{eqnarray}
which shows that the deviation in the reward caused by any unilateral action change of UAV equals to the change in the potential function. Thus, $\mathcal{G}_m$ is an exact potential game with the sum of the UAV rewards $\phi(a)$ serving as the potential function. The potential game has many nice properties \cite{4814554,8654694}. One of properties is that any potential game has at least one pure Nash equilibrium. This completes the proof.
\end{IEEEproof}
\begin{remark}
Theorem \ref{eq:theorem2} illuminates that every action profile $a^*$ maximizing the potential function is a pure Nash equilibrium, but it may be just a local optimal solution \cite{4814554}. Based on this, we can deduce that the global optimal solution and the pure Nash equilibrium have no one-to-one corresponding relations. Mathematically, let $\mathcal{A}^g$ be the action set of the global optimal solution and $\mathcal{A}^*$ be the action set of the pure Nash equilibrium, then the inclusion can be established, i.e., $\mathcal{A}^g\subset\mathcal{A}^*$.
A pure Nash equilibrium can be suboptimal, which makes the whole system sometimes just yield a local optimal solution. In other words, the pure Nash equilibrium is a necessary and not sufficient condition for the global optimal solution. Therefore, the proposed UAV channel capacity game can have more than one pure Nash equilibrium. One of the important tasks is to fully explore the environment in each time slot and achieve the pure Nash equilibrium with respect to the global optimal solution, namely the best Nash equilibrium.
\end{remark}
\section{Achieving Best Nash Equilibrium Using Decentralized Learning Algorithm}\label{sec:4}
For the proposed UAV channel capacity maximization game, one interesting property is the \textit{finite improvement property} \cite{revisting}, that if exactly one UAV is scheduled to change its strategy using a best reply process in each iteration, the value of the potential function can always increase. In this case, after a finite number of iterations, the strategy can converge to a pure Nash equilibrium. There are many learning algorithms that can make the potential function converge to a pure Nash equilibrium, such as the best response dynamic \cite{8654694}, the spatial adaptive play \cite{4814554,6086561}, the no-regret learning \cite{5506285} and the fictitious play \cite{revisting}. However, these algorithms may easily run into a local optimal solution \cite{4814554}. Therefore, we develop an effective learning algorithm that can achieve the best Nash equilibrium of the formulated game.
\subsection{Learning in game}
One important role for the developed learning algorithm is to guide UAVs to explore their optimal 3-D deployment using only the local information in real time and finally maximize the channel capacity of the whole system. Since the action space corresponds to the discrete position selection, when considering the collision avoidance problem of the UAV swarm, the UAVs cannot arbitrarily select actions from the action space. Specifically, if a candidate position that a UAV can select by performing action $a_m$ is occupied by any other UAVs, such specific action is restricted within the current time slot. We refer to such restricted action space as $\mathcal{A}_m^{\text{res}}(a_m^t)\subseteq \mathcal{A}_m$ where $a_m^t$ is the current action. Next, we give the following two properties on restricted action space.
\begin{property}
For any $m\in V$ and any action pair $a^0_m,a^s_m\in\mathcal{A}_m$, there exists a sequential actions $a^0_m\rightarrow\cdots\rightarrow a^{s-1}_{m}\rightarrow a^s_m$ satisfying $a^t_m\in \mathcal{A}_m^{\text{res}}(a_m^{t-1})$, $\forall t\in(1,\ldots,s)$.
\end{property}
\begin{property}
For any $m\in V$ and any action pair $a^0_m,a^1_m\in\mathcal{A}_m$, if $a^1_m\in \mathcal{A}_m^{\text{res}}(a_m^0)$, then, we can get $a^0_m\in \mathcal{A}_m^{\text{res}}(a_m^1)$.
\end{property}
\begin{remark}
Property 1 reveals that the action space available at the $m$-th UAV in time slot $t$ is a function of its action in previous time slot $t-1$ with restricted action space, and any action profile in $\mathcal{A}$ can be reached in finite time slots. Property 1 is known as reachability. Property 2 implies that each UAV can go back to its previous action. Property 2 is referred to as reversibility.
\end{remark}
The learning algorithm of the UAV channel capacity game is motivated by the binary Log-linear learning \cite{revisting,MJ1994}, which can handle game models with constrained action space.
In order to avoid running into a local Nash equilibrium, each UAV has to adequately explore the environment. Hence, we use a novel action choice heuristic based on the Boltzamnn exploration strategy, which states that each UAV chooses an action to perform in the next iteration of the game with a probability that is determined by the value of a temperature $T$. Specifically, in each time slot $t$, a UAV $m\in V$ is randomly chosen and allowed to change its action. Meanwhile, all the other UAVs still choose their previous actions, i.e., $a^t_{\mathcal{N}_m}=a^{t-1}_{\mathcal{N}_m}$. The probability of the position selection strategy of the $m$-th UAV can be given by
\begin{eqnarray}
\label{eq:Boltzamnn}
&P(a_m^t=a^{t-1}_m)=\frac{e^{\frac{1}{T}R_m(a^{t-1}_m,a^{t-1}_{\mathcal{N}_m})}}{e^{\frac{1}{T}R_m(a^{t-1}_m,a^{t-1}_{\mathcal{N}_m})}+e^{\frac{1}{T}R_m(\tilde{a}_m,a^{t-1}_{\mathcal{N}_m})}}\nonumber\\
&P(a_m^t=\tilde{a}_m)=\frac{e^{\frac{1}{T}R_m(\tilde{a}_m,a^{t-1}_{\mathcal{N}_m})}}{e^{\frac{1}{T}R_m(a^{t-1}_m,a^{t-1}_{\mathcal{N}_m})}+e^{\frac{1}{T}R_m(\tilde{a}_m,a^{t-1}_{\mathcal{N}_m})}}\nonumber\\
&P(a_m^t\neq a^{t-1}_m,\tilde{a}_m)=0,~~~~~~~~~~~~~~~~~~~~~~~~~~~~~
\end{eqnarray}
where $\tilde{a}_m$ is an action that is explored uniformly from the restricted action space $\mathcal{A}_m^{\text{res}}(a_m^{t-1})/\ a_m^{t-1}$, $P(a_m^t=a^{t-1}_m)$ is the probability that the $m$-th UAV remains on its previous action and $P(a_m^t=\tilde{a}_m)$ is the probability that the $m$-th UAV selects the action $\tilde{a}_m$.
\begin{remark}
When $T=0$, the position selection strategy boils down to an asynchronous best reply process. Specifically, in each time slot $t$, the $m$-th UAV selects the current action $a_m^t$ from $\tilde{a}_m\in\mathcal{A}_m^{\text{res}}(a_m^{t-1})$ or $a_m^{t-1}$ to maximize its reward function; at the same time, the other UAVs still choose their current actions, i.e., $a_m^t\in\{\tilde{a}_m,a_m^{t-1}\}:R_m(a_m^t,a^{t-1}_{N_m})=\max\{R_m(a^{t-1}_m,a^{t-1}_{N_m}),R_m(\tilde{a}_m,a^{t-1}_{N_m})\}$. When $T>0$, then each UAV has a chance to select locally a suboptimal action with a non-zero probability. In this way, the environment is fully explored, which can prevent the UAVs from converging to a local pure Nash equilibrium.
\end{remark}

\begin{algorithm}[!h]
\caption{Decentralized UAVs channel capacity learning}\label{alg:1}
\begin{algorithmic}[1]
\STATE \textbf{Initialization:} Set the iteration index $t=0$, let each UAV $m, m\in V$ randomly select a position $a_m^0\in \mathcal{A}_m$.
\FOR {$t=1,2,\ldots$}
\STATE Randomly choose at most one UAV, say $m$, according to an uniform distribution.
\STATE UAV $m$ communicates with its neighbours and constructs the local information $\mathcal{F}^t_m$, and then calculates its current reward $R_m(a^t_m,a^t_{\mathcal{N}_m})$.
\STATE UAV $m$ selects one exploration action $\tilde{a}_m$ from its restricted action space $\mathcal{A}_m^{\text{res}}(a_m^{t})\subseteq \mathcal{A}_m$ according to the following rule
\begin{eqnarray}
\label{eq:exploration}
P(a_m^{t+1})=\left\{
  \begin{array}{ll}
    \frac{\varepsilon_m}{|\mathcal{A}_m^{\text{res}}(a_m^t)|-1}, & \hbox{$a_m^{t+1}=\tilde{a}_m$} \\
    1-\varepsilon_m, & \hbox{$a_m^{t+1}=a_m^{t}$,}
  \end{array}
\right.
\end{eqnarray}
where $|\mathcal{A}_m^{\text{res}}(a_m^t)|$ represents the size of the restricted action space $\mathcal{A}_m^{\text{res}}(a_m^t)$, and $\varepsilon_m$ is referred to as the exploration rate.
\IF{$a_m^{t+1}=\tilde{a}_m$}
\STATE UAV $m$ calculates its explored reward $R_m(\tilde{a}_m,a^{t+1}_{\mathcal{N}_m})$ and updates the action selection according to (\ref{eq:Boltzamnn}). Meanwhile, all the other UAVs still choose their previous actions, i.e., $a^{t+1}_{\mathcal{N}_m}=a^t_{\mathcal{N}_m}$.
\ELSE
\STATE Return to step 2 and repeat.
\ENDIF
\ENDFOR
\end{algorithmic}
\end{algorithm}
The details of the developed algorithm are shown in Algorithm \ref{alg:1}. Obviously, such algorithm is decentralized since the learning process only depends on the local information exchanging among neighbors. Specifically, the algorithm can be divided into three stages, including the UAV selection stage, i.e., steps 3-4, the exploration stage, i.e., step 5, and the exploitation stage, i.e., steps 6-10. The whole procedure of the developed learning algorithm can be described as an interaction loop between the selected UAV and the environment, where the UAV chooses an action via the rule (\ref{eq:exploration}), then performs the action and gets the reward from the environment feedback, after which it updates the probability of the position selection strategy based on (\ref{eq:Boltzamnn}). The stop criterion of the algorithm can be met when the maximum number of iterations is reached or when the probability of the position selection strategy in (\ref{eq:Boltzamnn}) is asymptotically equal to 1, i.e., $P(a_m^t)\approx1, \forall m\in V$.
\subsection{Algorithmic analysis}
\subsubsection{Convergence analysis}
To guarantee the algorithm convergence, the probability of exploring a new action from the restricted action space should decrease with the iteration number and attain a very small positive value. More specifically, at the beginning of iterations, the exploration probability for a new action should take as large value as possible and keep decreasing with the number of iterations. Note that this probability declines faster in the early stage and then slowly converges to 0 in the later stage. Hence, we use a exponential function to characterize such behavior. The exploration rate in (\ref{eq:exploration}) is set to be $\varepsilon_m=e^{-\beta_m}$, where $\beta_m$ is the learning parameter linearly increasing with the iteration number.

\begin{theorem}
\label{eq:convergence}
If all the UAVs adhere to the proposed decentralized UAVs channel capacity learning algorithm, this algorithm asymptotically converges to the stochastically stable state for sufficiently large $\beta_m,\forall m\in V$, where $\beta_m=\ln\frac{1}{\varepsilon_m}$.
\end{theorem}
\begin{IEEEproof}
See Appendix \ref{app:1}.
\end{IEEEproof}
\begin{remark}
When the proposed learning algorithm converges, there still exists an extremely small exploration probability to keep exploring the restricted action space, just as the exploration and exploitation procedures in reinforcement learning. In this case, there is a probability $1-\Pi_{m\in V}(1-\varepsilon_m)$ that the whole system cannot converge to the stochastically stable state. However, the proposed learning algorithm is different from the reinforcement learning algorithm, i.e., Q-learning. One of the obvious difference is that the reinforcement learning is based on the long-term excepted reward
while our algorithm is based on the instantaneous reward and a potential function. For another, there are no assumption of the Markov decision process and update of the Q-function in our algorithm.
\end{remark}
\subsubsection{Effectiveness analysis} Here, we analyze that the proposed learning algorithm can effective address the rank coupling problem in the original optimization formulation (\ref{eq:optimal1}).
\begin{theorem}
\label{eq:effectiveness}
When the decentralized UAVs channel capacity learning algorithm converges, for all optimization sub-problems of all UAVs, the channel matrix rank $\mathbf{H}_m, \forall m\in V$ can be maximum, which can then maximize the rank of the channel matrix $\mathbf{H}$.
\end{theorem}
\begin{IEEEproof}
Suppose the channel matrix rank of the MIMO system is not maximum, then the channel matrix $\mathbf{H}$ has some column vectors $\mathbf{h}_i, \mathbf{h}_m, i,m\in V$ that are not linearly independent. In this case, there exists $\mathbf{h}_i=c\mathbf{h}_m, \exists i,m\in V, i\neq m$, and then $\text{rank}(\mathbf{H})<\min(M,N)$. On the other hand, for the $m$-th UAV, $\forall m\in V$, the channel matrix of each UAV can be written as  $\mathbf{H}_m=[\mathbf{h}_1,\ldots,\mathbf{h}_i,\mathbf{h}_m]$. By using the decentralized UAVs channel capacity learning algorithm, we can get $\text{rank}(\mathbf{H}_m)=\min(i+1,N)$ and $\mathbf{h}_i\neq c\mathbf{h}_m, \forall i,m\in V$. Therefore, the assumption cannot hold, and then the proof is completed.
\end{IEEEproof}
\begin{remark}
If the formulated graph $G=(V,E)$ for the inter-UAV communication is a complete graph, then, we have $\text{rank}(\mathbf{H}_m)=\text{rank}(\mathbf{H})=\min(M,N), \forall m\in V$. In this case, the information $\mathcal{F}^t_m = \mathcal{F}^t$ for any $m$ and $t$.
\end{remark}
\subsubsection{Complexity analysis}
The  computational complexity of the proposed learning algorithm includes the exploration update complexity and the iteration complexity. Regarding the former complexity, the algorithm needs a random number to choose one UAV $m$ with a computational complexity of $\mathcal{O}(1)$. Then, a computational complexity on the order of $\mathcal{O}(|\mathcal{N}_m|)$ and $\mathcal{O}(1)$ are required to calculate the reward and decide the position selection, respectively. The updating procedure involves $2$ exponents, $1$ addition and $4$ multiplications, thus the computational complexity is $\mathcal{O}(1)$. Therefore, the total computational complexity of the first one is on the order of $\mathcal{O}(|\mathcal{N}_m|)$. On the other hand, the iteration complexity of the algorithm is $\mathcal{O}(\bar{t})$, where $\bar{t}$ is the maximum number of iterations for the algorithm convergence. Overall, the total computational complexity is on the order of $\mathcal{O}(\bar{t}\cdot|\mathcal{N}_m|)$.
\begin{remark}
The computational complexity of the proposed learning algorithm is increased linearly with the number of neighbors and also increases linearly with the number of iterations. This observation indicates that the proposed learning algorithm has a feasible low computational complexity.
\end{remark}
\section{Simulation Results}\label{sec:5}
\begin{figure}[!ht]
 \centering
  \includegraphics[width=8.5cm]{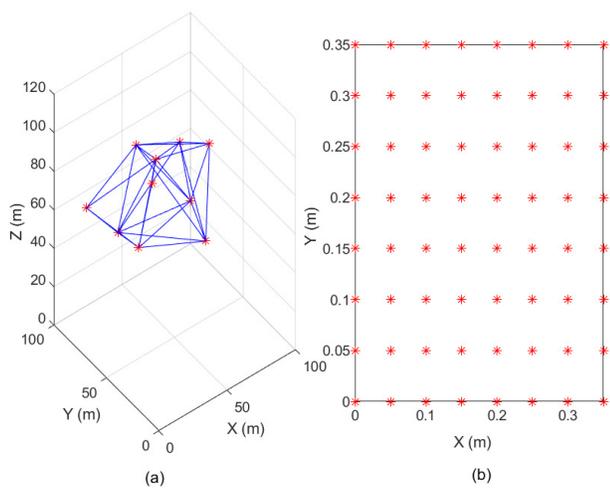}
  \caption{(a) The random deployment of UAVs in the initial stage and the neighbor layout of the UAVs; (b) The position of URA in the orthonormal coordinate system.}
  \label{Fig:UAVantennas}
\end{figure}

In this section, we evaluate the performance of the proposed decentralized UAVs channel capacity learning algorithm via simulations. In the simulations, the number of UAVs in the swarm is set to be 10 and the number of antennas in the URA is set to be 64 arranged as an 8$\times$8 array.
The initial deployments of both UAVs and the URA are shown in Fig. \ref{Fig:UAVantennas}. Specifically, we initially deploy the UAVs in a 100 m $\times$ 100 m $\times$ 120 m cubic area. The positions of UAVs are randomly selected, which follow a uniform distribution. The neighbors of each UAV are determined by the communication distance between the UAVs which is set to be 50 m. The antennas are placed in a 0.35 m $\times$ 0.35 m rectangular region with 5 cm spacing between each antenna.
For the air-to-ground LoS channel, we adopt the sub-6G frequency band and set the wavelength to be 1 cm. The initial learning parameter $\beta_m$ is set to be 0.01 and increases in each iteration with the step size 0.001.

\begin{figure}[!ht]
 \centering
  \includegraphics[width=8.5cm]{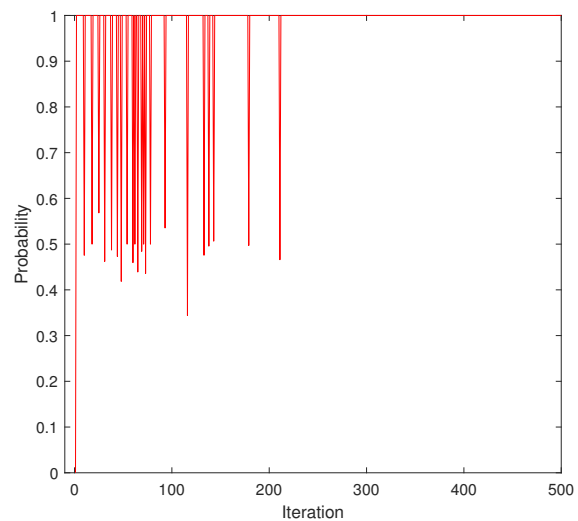}
  \caption{The probability of the position selection strategy against a different number of iterations.}
  \label{Fig:probability}
\end{figure}

\begin{figure}[!ht]
 \centering
  \includegraphics[width=8.5cm]{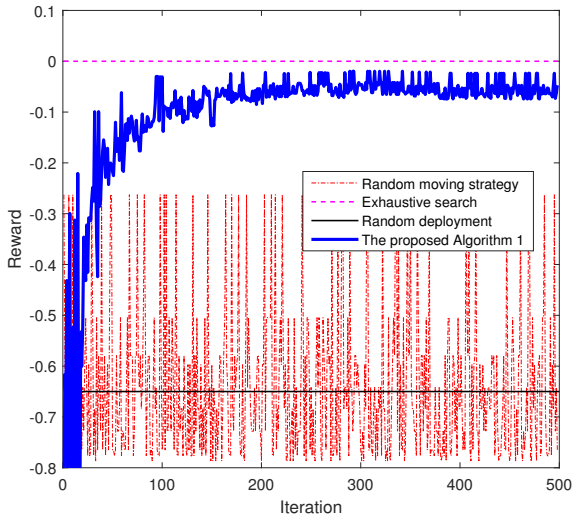}
  \caption{The reward of each UAV in 500 iterations with four different methods: the Algorithm 1, the random moving strategy, the exhaustive search and the random deployment \cite{kang20203d,8760424,8727504}.}
  \label{Fig:Reward}
\end{figure}

The probability of the position selection strategy with respect to the iteration number is shown in Fig. \ref{Fig:probability}. It can be observed that the developed learning algorithm prefers to explore the environment by choosing a new action rather than choosing the previous action, while it tends to select the previous action as the number of iterations increase. After 500 iterations, we can observe that the probability of the position selection can approximately converge to 1. In this case, all UAVs can realize the consensus control after 500 times interaction with the environment. It also reveals that the UAVs stop exploring a new position selection strategy at this time and the learning algorithm converges to a constant. It seems that such result is not well matched with the theoretical analysis, where in the learning algorithm there still exists an extremely small exploration probability to
keep exploring the restricted action space. The reason for this phenomenon is that the exploration rate $\varepsilon_m=e^{-\beta_m}$ keeps decreasing with the number of iterations and when $\beta_m$ is sufficiently large, the computer simulation program inherently sets the exploration rate to be 0, that is, $\lim_{\beta_m\rightarrow\infty}e^{-\beta_m}=0$. Therefore, this implies that it is initially important to select a suitable $\beta_m$ to learn the global optimal solution before $\beta_m$ becomes large enough and the simulation results return a zero value for the exploration rate. By using a suitable $\beta_m$ setting, the best Nash equilibrium can be achieved and the MIMO channel capacity can be maximized.
The performance of the developed learning algorithm is compared with benchmark methods in Fig. \ref{Fig:Reward}. Similar to references \cite{kang20203d,8760424,8727504}, the benchmark deployment strategies are: random moving, exhaustive search and random deployment. From this figure, we can see that the reward of the developed Algorithm 1 can approximately converge to -0.08 after 300 iterations, which is very close to 0. This suggests that the developed algorithm can successfully make the channel tend to be orthogonal, thereby maximizing the channel capacity. For the benchmark methods, the reward of the random moving strategy fluctuates between -0.8 and -0.27, and the reward of the random deployment is a constant around -0.65 since the position of the UAVs is unchanged over time. More specifically, the exhaustive search has the best performance of all the methods, however, this method needs the global information and takes a large amount of overhead, which is not suitable for our communication scenario. Therefore, according to the considered communication scenario, our developed learning algorithm outperforms the benchmark methods in terms of the orthogonality of the MIMO channel matrix.

\begin{figure}[!ht]
 \centering
  \includegraphics[width=8.5cm]{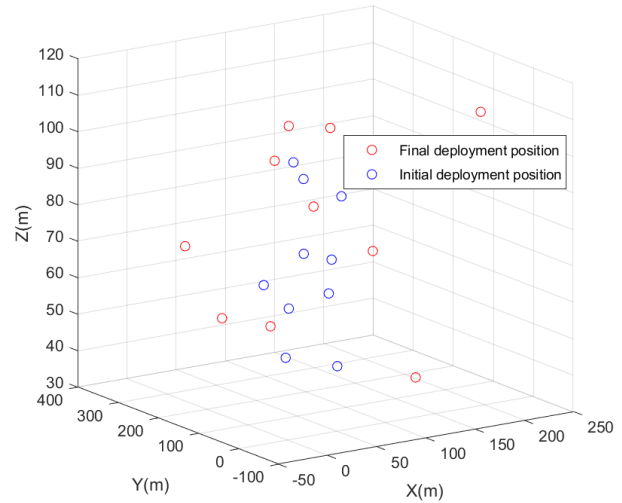}
  \caption{The final deployment position vs. the initial deployment position of the UAV swarm with 500 iterations.}
  \label{Fig:Fposition}
\end{figure}

Fig. \ref{Fig:probability} and Fig. \ref{Fig:Reward} are analyzed from the perspective of each individual UAV; in the following, the performance improvements are analyzed for the whole UAV swarm connected MIMO system. Fig. \ref{Fig:Fposition} shows the position of the UAV swarm in the 3-D deployment. For the UAV swarm, the blue circles represent the initial deployment positions and the red circles represent the final deployment positions after 500 iterations. We can observe that the final deployment positions can be more scattered than the initial deployment positions. The result shows that the developed learning algorithm tries to reduce the channel correlations by expanding the spacings among UAVs. It is easily understandable that when the spacings among UAVs are large, the distance difference between any two UAV and URA becomes significant, such that columns of the MIMO channel matrix can be linear independent.

\begin{figure}[!ht]
 \centering
  \includegraphics[width=8.5cm]{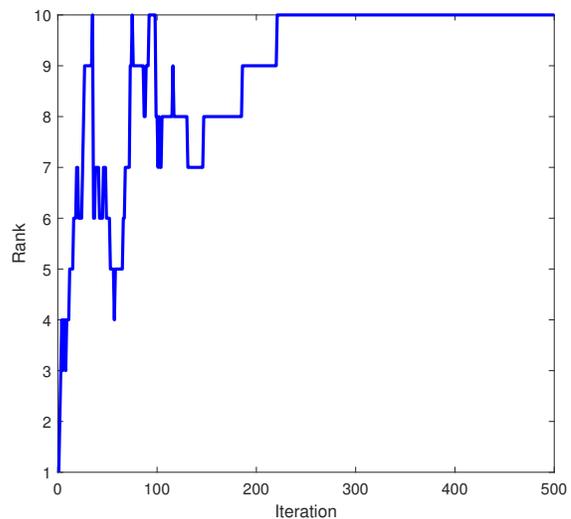}
  \caption{The rank of the MIMO channel matrix in 500 iterations with 10 UAVs and 64 URA antennas.}
  \label{Fig:rank1}
\end{figure}
\begin{figure}[!ht]
 \centering
  \includegraphics[width=8.5cm]{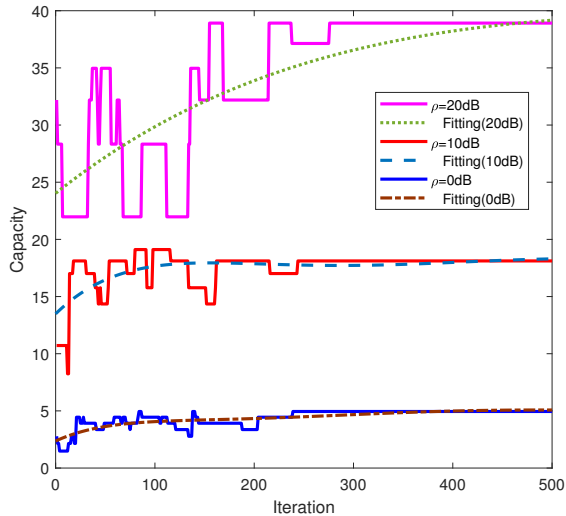}
  \caption{The channel capacity of the MIMO system in 500 iterations with different SNRs.}
  \label{Fig:capacity}
\end{figure}

In order to intuitively observe the change of the rank with respect to the iteration number, we now present Fig. \ref{Fig:rank1}. In this figure, we find that the rank of the MIMO channel matrix exhibits substantial fluctuations in the first 240 iterations and tends to be stable in the remaining iterations. After 500 iterations, the rank of the MIMO channel matrix is capable of reaching 10. In our simulation settings, there are 10 UAVs and 64 URA antennas; theoretically, the maximum rank of the MIMO channel matrix is $\min(10,64)=10$, which matches well with our simulation result. Overall, the rank of the MIMO channel matrix changes from 1 to 10 with the number of iterations, which shows a significant performance improvement for the MIMO channel capacity in the high SNR range.

In Fig. \ref{Fig:capacity}, we show the channel capacity of the MIMO system against the number of iterations. With fixed SNR, we can observe that the channel capacity increases in a fluctuating manner in the first few hundred iterations and steadily converges to a fixed value. For example, when the SNR$=$10 dB, the channel capacity increases in the first 255 iterations and converges to 18 bits/s/Hz at the 255-th iteration. When the SNR increases from 0 dB to 10 dB, the channel capacity has 13 bits/s/Hz improvement, while when the SNR increases from 10 dB to 20 dB, the channel capacity has 20 bits/s/Hz improvement. It suggests that it is more obvious to improve MIMO channel capacity by deploying UAV swarm with the increase of SNR. This finding matches well with the theoretical analysis in (\ref{eq:capcaity2}). In addition, the fitted channel capacity curves intuitively show the variation trends of the MIMO channel capacity with an increasing number of iterations. Overall, the result validates the effectiveness of the proposed learning algorithm in improving the channel capacity of the MIMO system.

\section{Conclusion}\label{sec:6}
We have investigated the 3-D deployment of UAVs to maximize the channel capacity of the MIMO system. Due to the difficulty for each UAV to obtain the global information of the networks, we have considered a decentralized control strategy and transformed the problem into several sub-problems, and then formulated a UAVs channel capacity maximization game to solve. The best Nash equilibrium was achieved based on the developed learning algorithm. Particularly, we provided the details of the algorithm, and then proved the convergence, the effectiveness and the computational complexity, respectively. Based on the proofs and the theoretical analysis, we found that each UAV can use only the local information from its neighbors to effectively maximize the channel capacity of the MIMO system. Simulation results validated the performance of the developed learning algorithm. Combining theory and simulations, we elucidated the importance of the exploration rate initialization in exploring the action space. It was also demonstrated that the nature of the developed learning algorithm is to reduce the channel correlations by expanding the spacings among UAVs. More importantly, we verified that the channel capacity of the MIMO system can achieve a high value and converge to this value via the proposed local information based optimal 3-D UAV swarm deployment.
\appendices
\section{Proof of Theorem 3}\label{app:1}
Since the procedure of the proposed learning algorithm is a Markovian process, the exact stationary distribution of states is hard to obtain \cite{revisting,6086561}. By using the theory of resistance tree, we study the stochastically stable states and then prove the convergence of the proposed learning algorithm.
\subsection{Brief review of resistance tree}
Let $P^0$ be the probability transition matrix for the best reply process over the state space $\mathcal{S}$ and the system state in time slot $t$ be $s^t=(a_m^t,a^t_{-m})$. We refer to $P^0$ as the unperturbed Markov process and $P^0_{s\rightarrow s'}$ as the probability of transition from state $s$ to $s'$. The probability transition matrix of the regular perturbed Markov process is defined by $P^\epsilon$ and the size of the perturbations can be indexed by $\epsilon>0$. The process $P^\epsilon$ has the following conditions:
\begin{enumerate}
  \item $P^\epsilon$ is aperiodic and irreducible.
  \item $P^\epsilon$ approaches $P^0$ at an exponentially smooth rate, i.e,
   \begin{eqnarray}
   \lim_{\epsilon\rightarrow0^+}P^\epsilon_{s\rightarrow s'}=P^0_{s\rightarrow s'}, \forall s,s'\in \mathcal{S}.
   \end{eqnarray}
  \item If $P^\epsilon_{s\rightarrow s'}$ for some $\epsilon>0$, there exists
   \begin{eqnarray}
\label{eq:0tolimit}
   0<\lim_{\epsilon\rightarrow0^+}\epsilon^{-\mathcal{R}(s\rightarrow s')}P^\epsilon_{s\rightarrow s'}<\infty,
   \end{eqnarray}
where $\mathcal{R}(s\rightarrow s')\in\mathbb{ R}^+$ is the resistance of the transition $s\rightarrow s'$.
\end{enumerate}
Construct a tree with $|\mathcal{S}|$ vertices with each state being a vertex. A tree $\mathcal{T}$ rooted at any vertex $v_s$ is a set of $|\mathcal{S}|-1$ directed edges such that there exists a unique directed path from every other state to $s$. The weight of a directed edge from vertex $v_s$ to $v_s'$ can be characterized by the resistance $\mathcal{R}(s\rightarrow s')$. The resistance of a rooted tree $\mathcal{T}$ is represented by the sum of the resistances of every connected vertices and the stochastic potential of state $s$ is defined to be the minimum resistance over all trees rooted at vertex $v_s$. The following lemma provides a criterion for determining the stochastically stable state of a regular perturbed Markov process.
\begin{lemma}
For each $\epsilon>0$, if $\mu^\epsilon$ is the unique stationary distribution of the regular perturbed Markov process $P^\epsilon$, then $\lim_{\epsilon\rightarrow0^+}$ exists and the limiting distribution $\mu^0$ is a stationary distribution of the unperturbed Markov process $P^0$. Moreover, the stochastically stable states are precisely those states with minimum stochastic potential \cite{revisting}.
\end{lemma}
\subsection{The asymptotic optimality of the proposed learning algorithm}
\begin{theorem}
\label{eq:theoremperturbed}
According to the procedure of the proposed learning algorithm, the position selection strategy is a regular perturbed Markov process, where the resistance of any feasible transition $s\rightarrow s'$ is given by
\begin{eqnarray}
\mathcal{R}(s\rightarrow s')=\max\{R_m(s),R_m(s')\}-R_m(s').
\end{eqnarray}
\end{theorem}
\begin{IEEEproof}
According to the procedure of the developed learning algorithm, the probability of transition from $s$ to $s'$ is
\begin{eqnarray}
\label{eq:resistance}
P_{s\rightarrow s'}^\epsilon=
\frac{U\varepsilon_m}{|\mathcal{A}_m^{\text{res}}(a_m)|-1}\cdot\frac{\epsilon^{-R_m(s')}}{\epsilon^{-R_m(s)}+\epsilon^{-R_m(s')}},
\end{eqnarray}
where $U=1/M$, $\epsilon=e^{-\frac{1}{T}}$. As the reward of the $m$-th UAV is decided by the action of its own and the actions of its neighbors, then, $P_{s\rightarrow s'}^\epsilon$ can be rewritten as
\begin{eqnarray}
\label{eq:resistance}
P_{s\rightarrow s'}^\epsilon=~~~~~~~~~~~~~~~~~~~~~~~~~~~~~~~~~~~~~~~~~~~~~~~~~~~~\nonumber\\
\frac{U\varepsilon_m}{|\mathcal{A}_m^{\text{res}}(a_m)|-1}\cdot\frac{\epsilon^{-R_m(\tilde{a}_m,a_{\mathcal{N}_m})}}{\epsilon^{-R_m(a_m,a_{\mathcal{N}_m})}+\epsilon^{-R_m(\tilde{a}_m,a_{\mathcal{N}_m})}}.
\end{eqnarray}
The first term represents the probability that an arbitrary UAV $m$ explores its action from $a_m$ to $\tilde{a}_m$. The second term is the probability that UAV $m$ is updated with the new action $\tilde{a}_m$. Next, the maximum reward of the $m$-th UAV for any two action profiles $(a_m,a_{-m})$ and $(\tilde{a}_m,a_{-m})$ is defined by
\begin{eqnarray}
B_m(s,s')=\max\{R_m(s),R_m(s')\}.
\end{eqnarray}
For the right-hand of (\ref{eq:resistance}), multiplying the numerator and denominator by $\epsilon^{B_m(s,s')}$, then, we have
\begin{eqnarray}
\label{eq:resistance1}
P_{s\rightarrow s'}^\epsilon=
\frac{U\varepsilon_m}{|\mathcal{A}_m^{\text{res}}(a_m)|-1}~~~~~~~~~~~~~~~~~~~~~~~~~~~~~~~~~~~\nonumber\\
\times\frac{\epsilon^{B_m(s,s')-R_m(\tilde{a}_m,a_{\mathcal{N}_m})}}{\epsilon^{B_m(s,s')-R_m(a_m,a_{\mathcal{N}_m})}+\epsilon^{B_m(s,s')-R_m(\tilde{a}_m,a_{\mathcal{N}_m})}}.
\end{eqnarray}
Dividing (\ref{eq:resistance1}) by $\epsilon^{B_m(s,s')-R_m(\tilde{a}_m,a_{\mathcal{N}_m})}$, we can get
\begin{eqnarray}
\label{eq:resistance1}
\frac{P_{s\rightarrow s'}^\epsilon}{\epsilon^{B_m(s,s')-R_m(\tilde{a}_m,a_{\mathcal{N}_m})}}=
\frac{U\varepsilon_m}{|\mathcal{A}_m^{\text{res}}(a_m)|-1}~~~~~~~~~~~~~~~~~\nonumber\\
\times\frac{1}{\epsilon^{B_m(s,s')-R_m(a_m,a_{\mathcal{N}_m})}+\epsilon^{B_m(s,s')-R_m(\tilde{a}_m,a_{\mathcal{N}_m})}}.
\end{eqnarray}
Based on the definition of $B_m(s,s')$, for each UAV $m, m\in V$,  we have
\begin{eqnarray}
B_m(s,s')-R_m(s')
=\left\{
            \begin{array}{ll}
              0, & \hbox{$R_m(s)\leq R_m(s')$} \\
              \mathbb{R}^+, & \hbox{$R_m(s)>R_m(s')$}.
            \end{array}
          \right.
\end{eqnarray}
When $\epsilon\rightarrow 0^+$, $\frac{1}{\epsilon^{B_m(s,s')-R_m(a_m,a_{\mathcal{N}_m})}+\epsilon^{B_m(s,s')-R_m(\tilde{a}_m,a_{\mathcal{N}_m})}}$ approaches 1, thus, for sufficiently large $\beta_m$, we can have
\begin{eqnarray}
\label{eq:big0}
\lim_{\epsilon\rightarrow 0^+}\frac{P_{s\rightarrow s'}^\epsilon}{\epsilon^{B_m(s,s')-R_m(\tilde{a}_m,a_{\mathcal{N}_m})}}=\frac{U\varepsilon_m}{|\mathcal{A}_m^{\text{res}}(a_m)|-1}>0.
\end{eqnarray}
Apparently, (\ref{eq:big0}) satisfies the following condition
\begin{eqnarray}
0<\lim_{\epsilon\rightarrow 0^+}\frac{P_{s\rightarrow s'}^\epsilon}{\epsilon^{B_m(s,s')-R_m(\tilde{a}_m,a_{\mathcal{N}_m})}}<\infty,\nonumber
\end{eqnarray}
which is consistent with the condition in (\ref{eq:0tolimit}). Therefore, the procedure of the developed learning algorithm is a regular perturbed Markov process, where the resistance of any feasible transition $s\rightarrow s'$ is given by
\begin{eqnarray}
\mathcal{R}(s\rightarrow s')=\max\{R_m(s),R_m(s')\}-R_m(s'),\nonumber
\end{eqnarray}
and the proof is completed.
\end{IEEEproof}
\begin{theorem}
\label{eq:theoremperturbed1}
For any potential game with a finite number of UAVs, if all UAVs adhere to the  developed learning algorithm, then the stochastically stable state is the set of potential maximizers.
\end{theorem}
\begin{IEEEproof}
The set of potential maximizers can be represented by $\mathcal{A}^*=\{a\in \mathcal{A}:\phi(a)=\max_{a^*\in \mathcal{A}}\phi(a^*)\}$ which is the global optimal
solution. From Theorem \ref{eq:theoremperturbed}, we know that the procedure of the developed
learning algorithm induces a regular perturbed Markov process. We assume that there is a minimum resistance tree rooted at $a$ that does not maximize the potential function. Then, for the rooted tree $\mathcal{T}$, there exists a path $\mathcal{P}$ from action $a^*$ to $a$ of the form
\begin{eqnarray}
\mathcal{P}=\{a^*\rightarrow\cdots\rightarrow a'\rightarrow\cdots\rightarrow a\}.
\end{eqnarray}
Based on the reversibility in Property 2, the reverse path $\overleftarrow{\mathcal{P}}$ that goes from $a$ to $a^*$ can be given by
\begin{eqnarray}
\overleftarrow{\mathcal{P}}=\{a\rightarrow\cdots\rightarrow a'\rightarrow\cdots\rightarrow a^*\}.
\end{eqnarray}
By adding the edges of $\overleftarrow{\mathcal{P}}$ to $\mathcal{T}$ and removing the redundant edges $\mathcal{P}$, we get a new tree $\mathcal{T}'$ rooted at $a^*$. The resistance of the new tree is
\begin{eqnarray}
\mathcal{R}(\mathcal{T}')=\mathcal{R}(\mathcal{T})+\mathcal{R}(\overleftarrow{\mathcal{P}})-\mathcal{R}(\mathcal{P})\nonumber\\
=\mathcal{R}(\mathcal{T})+\phi(a)-\phi(a^*)~~\nonumber\\
<\mathcal{R}(\mathcal{T}).~~~~~~~~~~~~~~~~~~~~
\end{eqnarray}
Hence, the minimum resistance tree can maximize the potential function. Repeating the above analysis for all the action profiles that can maximize the potential function, we find they have the same stochastic potential. Then, the stochastically stable state is the set of potential maximizers, and the proof is completed.
\end{IEEEproof}
Note that, combining Theorem \ref{eq:theoremperturbed} and Theorem \ref{eq:theoremperturbed1}, we can get the convergence of the developed learning algorithm. Then, the proof of Theorem \ref{eq:convergence} is completed.



\ifCLASSOPTIONcaptionsoff
  \newpage
\fi



\bibliographystyle{IEEEtran}
\bibliography{IEEEabrv,sigproc} 
%

%

\begin{IEEEbiography}[{\includegraphics[width=1in,height=1.25in,clip,keepaspectratio]{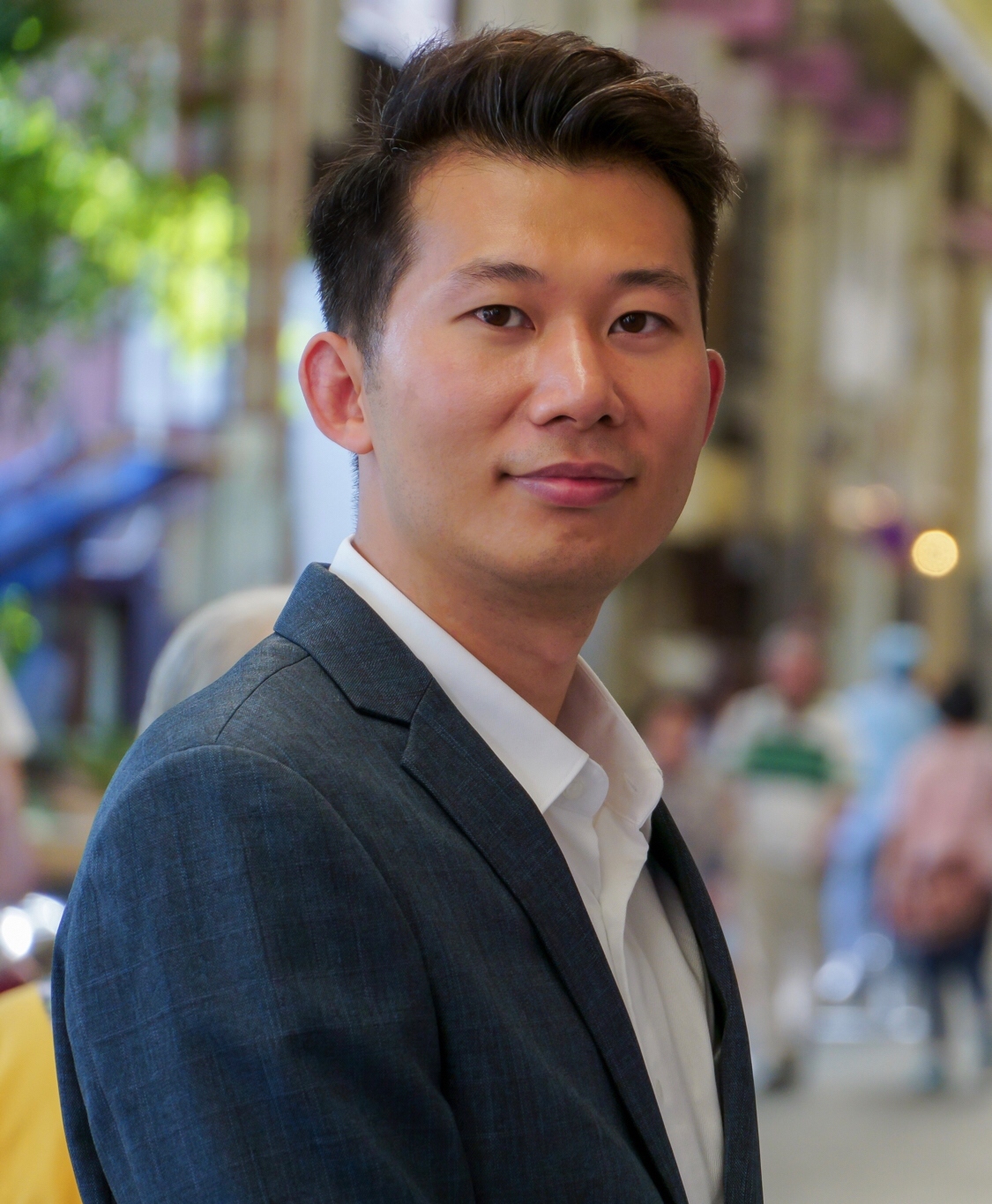}}]{Ning Gao}

received the Ph.D. degree in information and communications
engineering from the Beijing University
of Posts and Telecommunications, Beijing, China, in
2019. From 2017 to 2018, he was a Visiting Ph.D. Student with the School of Computing and
Communications, Lancaster University, Lancaster,
U.K. He is currently a Research Fellow with the National Mobile Communications Research Laboratory, Southeast University. His research interests include wireless eavesdropping and spoofing, intelligent communications and UAV communications. Dr. Gao was a recipient of the 2020 Postdoctoral Innovative Talents Program of China.
\end{IEEEbiography}
\begin{IEEEbiography}[{\includegraphics[width=1in,height=1.25in,clip,keepaspectratio]{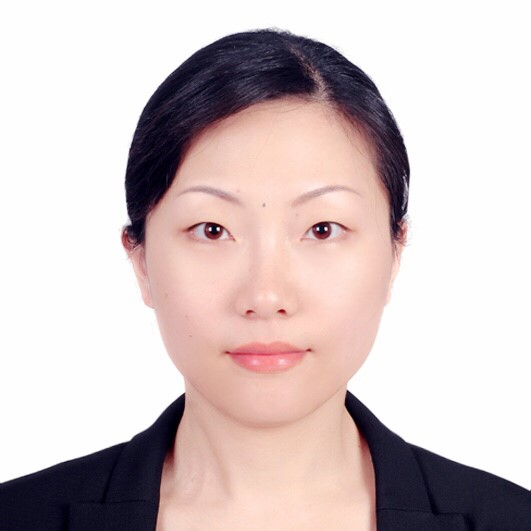}}]{Xiao Li}
(S’06-M’10) received the Ph.D. degree in communication and information systems from Southeast University, Nanjing, China, in 2010.
She then joined the School of Information Science and Engineering, Southeast University, where she has been a Professor in information systems and communications since July 2020. From January 2013 to January 2014, she was a Postdoctoral Fellow at The University of Texas at Austin, Austin, TX, USA. Her current research interests include massive MIMO, Reconfigurable intelligent surface assisted communications, and intelligent communications.
Dr. Li was a recipient of the 2013 National Excellent Doctoral Dissertation of China for her Ph.D. dissertation. She serves as an Associate Editor for the \textsc{IEEE Wireless Communications Letters} and \textsc{Electronics Letters}.

\end{IEEEbiography}
\begin{IEEEbiography}[{\includegraphics[width=1in,height=1.25in,clip,keepaspectratio]{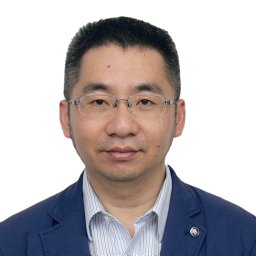}}]{Shi Jin}

(S'06-M'07-SM'17) received the B.S. degree in communications engineering from the Guilin University of Electronic Technology, Guilin, China, in 1996, the M.S. degree from the Nanjing University of Posts and Telecommunications, Nanjing, China, in 2003, and the Ph.D. degree in communications and information systems from Southeast University, Nanjing, China, in 2007. From June 2007 to October 2009, he was a Research Fellow with the Adastral Park Research Campus, University College London, London, U.K. He is currently with the Faculty of the National Mobile Communications Research Laboratory, Southeast University. His research interests include space–time wireless communications, random matrix theory, and information theory.
 He serves as an Associate Editor
for the  \textsc{IEEE Transactions on Wireless Communications}, and  \textsc{IEEE Communications Letters}, and  \textsc{IET Communications}. Dr. Jin and his co-authors have been awarded the 2011 IEEE Communications Society Stephen O. Rice Prize Paper Award in the field of communication theory and a 2010 Young Author Best Paper Award by the IEEE Signal Processing Society.
\end{IEEEbiography}

\begin{IEEEbiography}[{\includegraphics[width=1in,height=1.25in,clip,keepaspectratio]{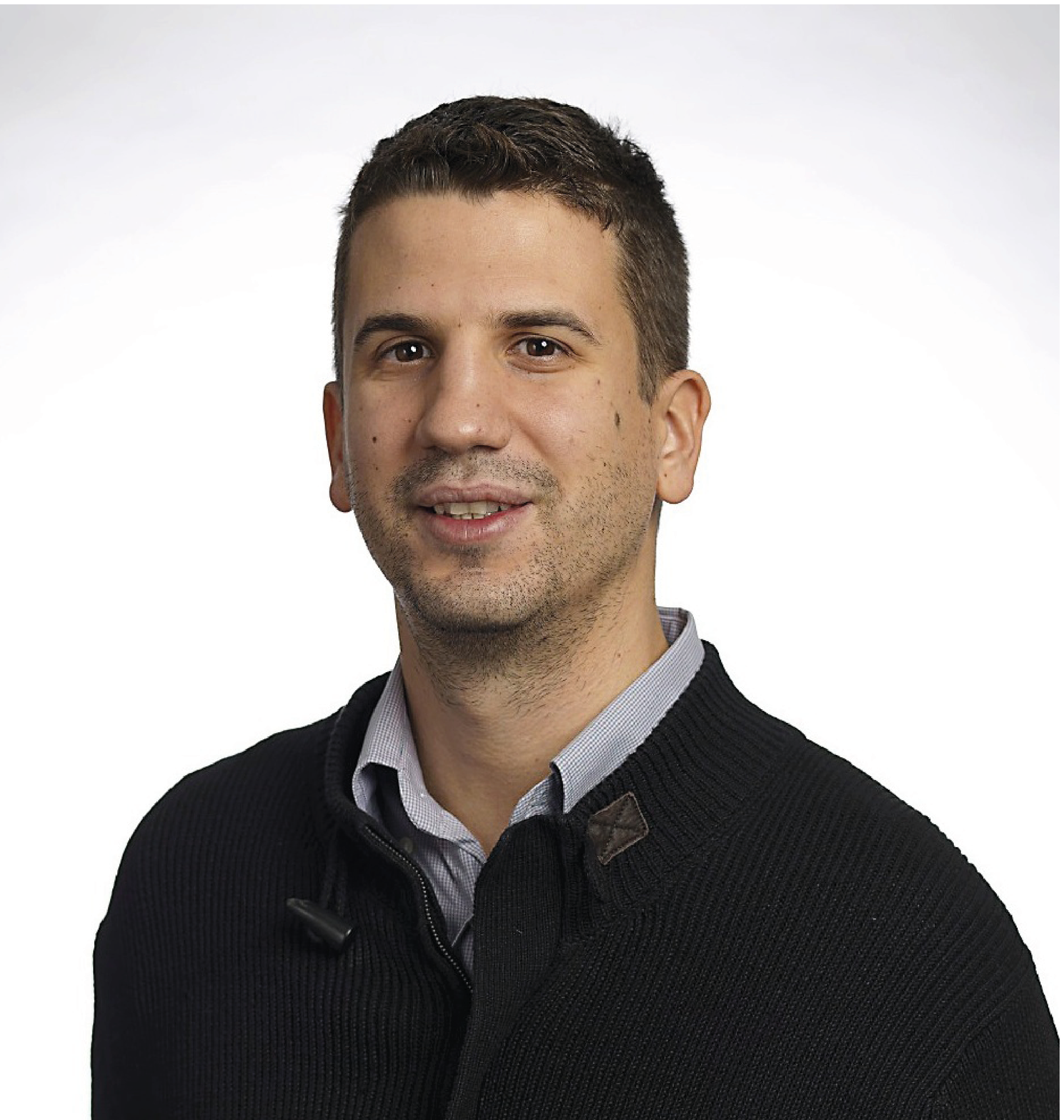}}]{Michail Matthaiou}
(S'05--M'08--SM'13) was born in Thessaloniki, Greece in 1981. He obtained the Diploma degree (5 years) in Electrical and Computer Engineering from the Aristotle University of Thessaloniki, Greece in 2004. He then received the M.Sc. (with distinction) in Communication Systems and Signal Processing from the University of Bristol, U.K. and Ph.D. degrees from the University of Edinburgh, U.K. in 2005 and 2008, respectively. From September 2008 through May 2010, he was with the Institute for Circuit Theory and Signal Processing, Munich University of Technology (TUM), Germany working as a Postdoctoral Research Associate. He is currently a Professor of Communications Engineering and Signal Processing and Deputy Director of the Centre for Wireless Innovation (CWI) at Queen’s University Belfast, U.K. after holding an Assistant Professor position at Chalmers University of Technology, Sweden. His research interests span signal processing for wireless communications, beyond massive MIMO, intelligent reflecting surfaces, mm-wave/THz systems and deep learning for communications.

Dr. Matthaiou and his coauthors received the IEEE Communications Society (ComSoc) Leonard G. Abraham Prize in 2017. He currently holds the ERC
Consolidator Grant BEATRICE (2021-2026) focused on the interface between information and electromagnetic theories. He was awarded the prestigious 2018/2019 Royal Academy of Engineering/The Leverhulme Trust Senior Research Fellowship and also received the 2019 EURASIP Early Career Award. His team was also the Grand Winner of the 2019 Mobile World Congress Challenge. He was the recipient of the 2011 IEEE ComSoc Best Young Researcher Award for the Europe, Middle East and Africa Region and a co-recipient of the 2006 IEEE Communications Chapter Project Prize for the best M.Sc. dissertation in the area of communications. He has co-authored papers that received best paper awards at the 2018 IEEE WCSP and 2014 IEEE ICC and was an Exemplary Reviewer for \textsc{IEEE Communications Letters} for 2010. In 2014, he received the Research Fund for International Young Scientists from the National Natural Science Foundation of China. He is currently the Editor-in-Chief of Elsevier Physical Communication, a Senior Editor for \textsc{IEEE Wireless Communications Letters} and an Associate Editor for the \textsc{IEEE JSAC Series on Machine Learning for Communications and Networks}.
\end{IEEEbiography}






\end{document}